%% file: petrinets.tex
\documentclass[submission]{eptcs}
\providecommand{\event}{FESCA 2013} 
\usepackage{times}
\usepackage{subfig}
\usepackage[boxed,lined]{algorithm2e}
\usepackage{xspace}
\usepackage{floatrow} 

\input{defaultCmds}

\newcommand{\timeoutend}{\ensuremath{\bar{t}}}
\newcommand{\timeoutbegin}{\ensuremath{\underline{t}}}
\newcommand{\InfArrow}{\ensuremath{\xrightarrow{[0,\infty)}}}
\newcommand{\pnarc}[1]{\ensuremath{\xrightarrow{#1}}}
\newcommand{\emptyarc}{\pnarc{\phantom{[t, t)}}}
\newcommand{\tarc}[1]{\ensuremath{\xrightarrow{#1\,\,\,} \hspace{-9px} \blacklozenge}}
\newcommand{\emptytarc}{\tarc{\phantom{[t, t)}}}
\newcommand{\InfTarc}{\tarc{[0,\infty)}}

\newcommand{\sd}{sequence diagram}

\input{tapn_tiks}

\title{Sequence Diagram Test Case Specification and\\ Virtual Integration Analysis using Timed-Arc Petri Nets}

\author{Sven Sieverding
\institute{OFFIS\\ Oldenburg, Germany}
\email{sieverding@offis.de}
\and
Christian Ellen
\institute{OFFIS\\ Oldenburg, Germany}
\email{ellen@offis.de}
\and
Peter Battram
\institute{OFFIS\\ Oldenburg, Germany}
\email{battram@offis.de}
}
\def\titlerunning{TCSD Specification and Virtual Integration Analysis using TAPN}
\def\authorrunning{S. Sieverding, C. Ellen \& P. Battram}
\begin{document}

\maketitle

\begin{abstract}
In this paper, we formally define Test Case Sequence Diagrams (TCSD) as an easy-to-use means to specify test cases for components including timing constraints.
These test cases are modeled using the UML2 syntax and can be specified by standard UML-modeling-tools.
In a component-based design an early identification of errors can be achieved by a virtual integration of components before the actual system is build.
We define such a procedure which integrates the individual test cases of the components according to the interconnections of a given architecture and checks if all specified communication sequences are consistent. 
Therefore, we formally define the transformation of TCSD into timed-arc Petri nets and a process for the combination of these nets.
The applicability of our approach is demonstrated on an avionic use case from the ARP4761 standard.

\end{abstract}

\input{introduction}

\input{methods}

\input{translation}

\input{evaluation}

\input{conclusion}

\paragraph{Acknowledgments}
The research leading to these results has received funding from the ARTEMIS Joint Undertaking under grant agreement $\text{n}^\circ 269335$.
It was also partially funded by the German Federal Ministry of Education and Research (BMBF), grant "SPES~XT, 01IS12005M" .


\bibliographystyle{eptcs}
\bibliography{references}

\end{document}

%% file: defaultCmds.tex
\usepackage{dsfont}
\usepackage{graphicx}
\usepackage{dcolumn}
\usepackage{bm}
\usepackage{todonotes}
\usepackage{hyperref}

\usepackage{amsmath}
\usepackage{amssymb}
\usepackage{amsthm}
\usepackage{caption}

\newcommand{\R}{\ensuremath{\mathds{R}}}
\newcommand{\Rposo}{\ensuremath{\R^+_0}}
\newcommand{\N}{\ensuremath{\mathbb{N}}}
\newcommand{\Ical}{\ensuremath{\mathcal{I}}}

\newcommand{\Tcal}{\ensuremath{\mathcal{T}}}
\newcommand{\Ccal}{\ensuremath{\mathcal{C}}}

\usepackage{rotating}

\newtheoremstyle{slplain}
  {0.5\baselineskip\@plus.2\baselineskip\@minus.2\baselineskip}
  {.5\baselineskip\@plus.2\baselineskip\@minus.2\baselineskip}
  {\itshape}
  {}
  {\bfseries}
  {}
  {\newline }
  {}

\theoremstyle{slplain}
\newtheorem{thm:def}{Definition}
\newtheorem{thm:rule}{Translation Rule}

\newcommand{\eg}{{\em e.g.,\,}}

\newcommand{\st}{{\em s.t.\,}}



%% file: tapn_tiks.tex
\usepackage{tikz}
\usetikzlibrary{arrows,shapes,snakes,automata,backgrounds,petri}

\tikzstyle{place}=[circle,thick,draw=blue!75,fill=blue!20,minimum size=7mm]
  \tikzstyle{red place}=[place,draw=red!75,fill=red!20]
  \tikzstyle{transition}=[rectangle,thick,draw=black!90,fill=black!70,minimum height=9mm, minimum width=4mm]
  \tikzstyle{transitionh}=[rectangle,thick,draw=black!90,fill=black!70,minimum height=4mm, minimum width=9mm]
  \tikzstyle{dots}=[rectangle,thick,draw=blue!90,dashed,fill=black!20,minimum height=10mm, minimum width=8mm]
  \tikzstyle{line} = [draw, -latex]
  \tikzstyle{plus}=[place,draw=red!75,fill=red!20, dashed, minimum size=10mm]

%% file: introduction.tex
\section{Introduction}

Testing is an important activity in modern embedded systems development
processes, \eg ISO 26262 \cite{ISO26262}, SAE ARP4761 \cite{ARP47611996}.
It comprises different level of granularity.
In case of a component-based design \cite{Damm2011} the unit-tests have to
validate that each individual component fulfills its requirements.
The integration-tests deal with the problems that arise through the
combination of multiple components and have to ensure their correct
interaction.
On the highest level, the complete system has to be validated.

In this paper, we focus on two testing aspects. First, an easy-to-use
specification method for unit-tests using UML2 syntax \cite{uml2}. Second,
we define an early virtual integration analysis on the basis of these test
cases.
For the test case specification, we introduce the concept of test case
sequence diagrams (TCSD) as an extension of UML2 sequence diagrams. This
extension allows test engineers to annotate timing constraints for messages
in the test case specification. Out of these test cases we are then able to generate
a formal analysis model to perform the virtual integration analysis.

The main idea of this approach is to interpret the successfully executed
unit-test cases as contract specification for the component. The pair of
input and expected result defines the assumption and the promise on the
connected input ports and output ports respectively. On the basis of these
specifications, we are analysing if test cases for different components of
the same system contradict each other regarding timing behavior or the
ordering of messages.

Our main contribution is the formalisation of sequence diagram-based test
cases and the virtual integration analysis of the test cases.
This approach is demonstrated on a well-known aerospace example from the
ARP4761, the braking system control unit (BSCU).
The components of the BSCU are implemented using Matlab/Simulink and
\sd -based test cases and the system architecture of the BSCU are modeled
using IBM Rational Rhapsody. These test cases
are translated into timed-arc Petri nets (TAPN) \cite{Hanisch1993}, which are
modeled and analyzed using the tool TAPAAL \cite{Byg2009}. For the virtual
integration, we developed a prototype which is able to export the necessary
information from Rhapsody. It also translates the exported TCSD into
timed-arc Petri nets, which can then be analyzed by TAPAAL.

The structure of this paper is as follows:
The formal definition of \sd s, test case \sd s, and Petri nets is
described in section \ref{sec:formalDefinition}. Section
\ref{sec:virtualIntegration} presents the virtual integration process,
including the translation mechanism of TCSD to TAPN. The demonstration of
our analysis is evaluated in section \ref{sec:evaluation} and in section
\ref{sec:conclusion} a conclusion as well as an outlook for our future work
is given.

\subsection{Related work}
There is a huge amount of publications available dealing with UML \cite{uml2} models for test case
specification. For example Sokenou \cite{Sokenou2006} identified the need to include
\sd s and state diagrams, which are usually created in the early stages of
the development, into his test sequence generation method. Also Linzhang
\cite{Linzhang2004} described the UML models as a natural source for test
case generation, since this semi-formal modelling language is commonly used.  

The formalization
of \sd s has been analyzed in \cite{Li2004,Micskei2010,Eichner2005} as well
as their transformation into other specifications. For example Bowles
\cite{Bowles2010} formalized \sd s and translated \sd s into coloured Petri
nets.
In this paper, we use his formalism for \sd s but translate them into
timed-arc Petri nets (TAPN), because of the annotated timing information.
We chose Petri nets over timed automata \cite{Srba2005} because of the
simplicity to compose different Petri nets in parallel
\cite{srba2008comparing}. 

The background of this work is based-on prior work 
\cite{Sieverding2011} in which the idea of sequence diagram-based test case
specification is introduced as well as the concept for consistency analysis.
This work elaborates on these ideas and creates a formal model for the test
case sequence diagrams.
We also define the virtual integration analysis on the formal model of TCSDs.

%% file: methods.tex
\section{Formal Definitions}
\label{sec:formalDefinition}
This section introduces the formal models for our virtual integration, namely sequence diagrams, timed-arc Petri nets, and test case sequence diagrams.

The basic models for UML sequence diagrams and timed-arc Petri nets are both based-on existing formal models, whereas test case sequence diagrams are a new extension to sequence diagrams specifically designed to suit the specification needs of timed behavior of a system under test.

\subsection{UML Sequence Diagrams}
\label{ssec:sd}
\begin{figure}
\floatbox[{\capbeside\thisfloatsetup{capbesideposition={right,top},capbesidewidth=0.4\textwidth}}]{figure}[\FBwidth]
{\caption{The figure shows examples of the structural elements of test case sequence diagrams and their different types of events.\\ 
For example: 
The box labeled with SUT is the instance line of the system under test, which interacts with the test components 1 \& 2;
the arrow labeled with $l_1$ is a message sent by event $e_{11}$ and received by $e_{21}$; 
the horizontal line with events $e_{12},e_{22},$ and $e_{31}$ is a partition line which models a timing constraint of $10ms$; 
the box labeled with \texttt{par} is a fragment in which the events of the two operands, seperated by the dashed line, are executed in parallel; 
and the arrow between event $e_{25}$ and $e_{29}$ is a timeout operator, which constraints the time between the occurrences of these events to $5ms$.
}\label{fig:example-sd}}
{\includegraphics[width=0.5\textwidth]{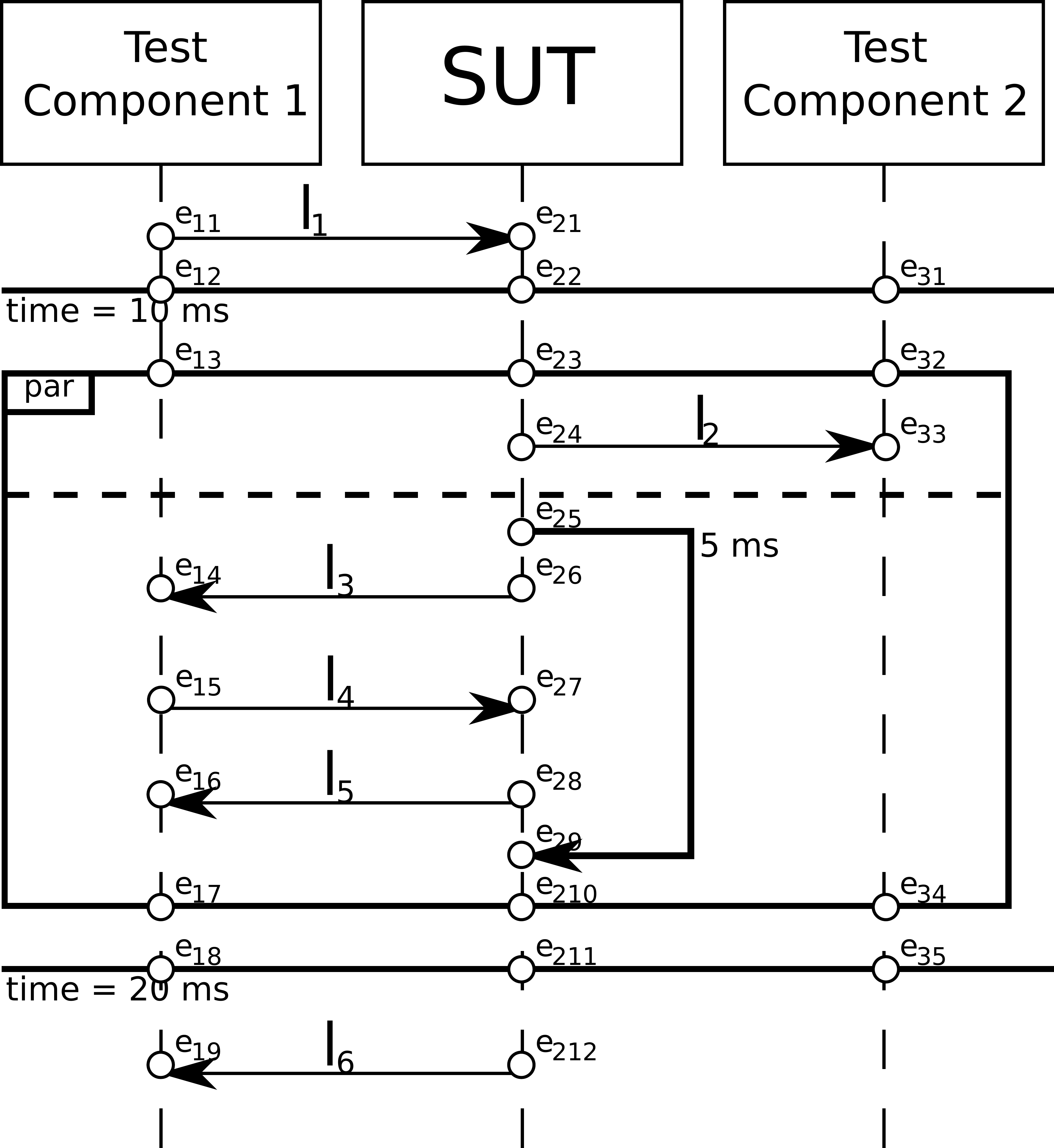}}

\end{figure}

The semantics of sequence diagrams, we use in this paper, are based-on the informal specification in the UML 2.0 superstructure \cite{uml2} and the formal semantics defined in Bowles and Meedeniya \cite{Bowles2010}. 
In addition, we define a slightly adapted version of these semantics, named \emph{test case sequence diagram} (TCSD), to deal with diagrams specifically designed for test cases. 
Figure \ref{fig:example-sd} shows an example of a TCSD and introduces the basic idea of the formalization, which is based-on the different types of events on the instance lines.

\begin{thm:def}[Sequence Diagram (based-on \cite{Bowles2010})]
\label{def:sd}
A sequence diagram (SD) is a tuple $D=(d,I,E,<,\Sigma_{msg},M,F,X,Exp)$ where:
\begin{itemize}
	\item $d\in \Sigma_{name}$ is the name of the diagram and $\Sigma_{name}$ the set of all diagram names;
	\item $I$ is a finite set of object instances (lifelines);

	\item $E=\bigcup_{i\in I}E_i$ is a set of events for lifeline $i$, \st $\forall i,j \in I: E_i \cap E_j = \emptyset$;

	\item $<$ is a set of partial orders which defines for instance line $i\in I$ a set: $<_i\subseteq E_i \times E_i$;

	\item $\Sigma_{msg}$ is a finite set of message labels $l$;

	\item $M$ is a set of messages $M \subseteq E \times \Sigma_{msg} \times E$, \st for every $m_1,m_2 \in M$ with $m_1=(e_{11},l_1,e_{12})$ and $m_2=(e_{21},l_2,e_{22})$: $m_1 \neq m_2 \implies e_{11} \neq e_{12} \neq e_{21} \neq e_{22} $;
	\item $F$ is a set of interaction fragments for which the functions $op, ev, sub$ are defined as:
	\begin{itemize}
	\item $op: F \rightarrow \Omega \times \N$ associates an operator $\Omega \in \lbrace \texttt{strict, par, opt, alt, loop} \rbrace$ and the number of operands to a fragment;
	\item $ev: F \times \N \rightarrow 2^E$ associates a set of events to a pair $(id, n)$ of a fragment $id\in F$ and an operand index number $n$; 
	\item $sub: F \times \N \rightarrow 2^F$ associates a set of nested fragments to a parent fragment and an operand index number; 
\end{itemize} 
\item $X = \lbrace X_i\rbrace_{i\in I}$ a set of local variables indexed by object instances $i \in I$;
\item $Exp$ is a set of expressions, where each expression is associated as a guard to a message or a fragment using the function $guard: M \cup F \rightarrow Exp$
\end{itemize}
\end{thm:def}

\noindent
A sequence diagram as defined in Definition \ref{def:sd} has a name $d$, which is usually used to identify a diagram in a UML modeling tool and a set of object instances $I$. Each instance $i\in I$ describes the behavior of one object in the diagram, which is defined by the events $E_i$ on the lifeline. All events within the diagram are partially ordered by a relation $<_i$. Only a partial ordering is possible, because the events in operands of fragment blocks like \texttt{par} cannot be ordered.
An event of a sequence diagram can either be part of a message (send event/receive event) or mark the borders of an interaction fragment (enter event/exit event). 
The interactions of a sequence diagram are defined by its transitions (messages) and the different kind of fragments. Each message $m$ is a tuple $m=(e_1,l_1,e_2)$ where $e_1 \in E_i$ is a send event, $l_1$ is a message label of the labeling alphabet $\Sigma_{msg}$, and $e_2 \in E_j$ is a receive event. Both events must be different ($e_1 \neq e_2$). In addition, the events must be ordered ($e_1 < e_2$) if they are part of the same instance line ($i=j$).

Fragments are regions within a sequence diagram with a specific semantic defined by the operator $\Omega$ of the fragment. 
A fragment spans over a subset of all instance lines of the diagram and every instance part of the fragment has dedicated \texttt{enter} and \texttt{exit} events which signal the fragments boundaries with respect to this instance.
The operators \texttt{strict, par, alt,opt,} and \texttt{loop} are relevant for our virtual integration scenario. The behavior of \texttt{strict} is the default behavior and requires that the events on the lifeline must occur in the specified order (according to the $<_i$-relation). A \texttt{par} fragment has at least two operands. Starting with the \texttt{enter} event, the event sequences of all operands are executed in parallel. The \texttt{exit} event of the \texttt{par} fragment is reached when all operands have reached this event following the $<_i$-relation. In contrast to the parallel execution of the \texttt{par} fragment, the operands of the \texttt{alt} fragment (at least two) represent exclusive alternative event sequences. The optional behavior of the \texttt{opt} fragment can be considered as a special case of an \texttt{alt} fragment, with one operand and an implicit empty sequence as second alternative. The \texttt{loop} fragment has exactly one operand. Its sequence of events is repeated as often as stated in the expression of its guard $Exp(f_{loop})$.

For each fragment three functions $op$, $ev$, and $sub$ are defined: 
\begin{itemize}
	\item[$op:$] The $op$ function assigns to each fragment an operator and the number of operands within the fragment. Each operator of a fragment requires a specific (minimum) number of operands. The \texttt{strict}, \texttt{loop}, and \texttt{opt} for example require exactly one operand, while the \texttt{par} and \texttt{alt} operators require at least two operands. 

	\item[$ev:$] The $ev$ function defines which elements are part of an operand. Therefore, it maps a tuple $(f,n)$ of a fragment $f$ and an operand  number $n$ to a set of events.
	
	\item[$sub:$] In sequence diagrams fragments may be nested. The function $sub$ describes this hierarchy by mapping a tuple $(f,n)$ to the set of the directly nested fragments within the $n$th fragment and therefore establishes a parent--child relation.
\end{itemize}

\noindent
To avoid ill-formed fragment and event hierarchies, all fragments of a sequence diagram must fulfill a set of consistency properties, \st for any fragments $f_1, f_2 \in F$ with $f_1 \neq f_2$, $op(f_1) = (o_1,n_1)$ and $op(f_2) = (o_2,n_2)$ the following properties must hold: 
\begin{enumerate}
	\item no self--nesting: $f_1 \notin sub(f_1,x)$ for any $x\leq n_1$
	\item no shared events (except if nested): $\forall x_1 \leq n_1, x_2 \leq n_2: f_1 \notin sub(f_2,x_1) \wedge f_2 \notin sub(f_1,x_2) \implies ev(f_1,x_1) \cap ev(f_2,x_2) = \emptyset$
	\item containment of events: $f_2 \in sub(f_1,x_1) \implies ev(f_1,x_1) \supseteq \bigcup_{n\leq n_2} ev(f_2,n)$
\end{enumerate}

\noindent
The first property avoids that any fragment may be nested in itself. The second property requires that any two disjunct fragments must not share events. The third property requires that if two fragments are nested the parent fragment must contain all events of its child fragments.

A sequence diagram can also contain a number of variables $X$ and expressions $Exp$ which are used on messages and fragments as guards. Apart from constant expressions on \texttt{loop} fragments, the variable and expression concepts are not relevant for our virtual integration analysis and are therefore ignored in the rest of this paper. 

\subsection{Test Case Sequence Diagrams}
\noindent
In our approach, we extend Definition \ref{def:sd} to \emph{test case sequence diagrams} (see Definition \ref{def:tcsd}), which are able to model the timed behavior of a test case. In a TCSD the object instances have two different roles. One instance represents the component of the system under test (SUT) for which the test case is defined. All other object instances of the diagram represent test components. Test components are abstract components which will (in most cases) not appear within the real system architecture. Their only purpose is to provide input to and receive output from the ports of the SUT. 

Therefore, we limit the set of events in messages, such that either the sending or the receiving event must be part of the $sut$ object instance and the other event must be part of a test component. Self loops on the SUT are not allowed because the idea of the test components is to make the communication with the SUT visible, such that messages to the SUT can be interpreted as part of an input test vector and messages received from the SUT as part of the expected output.

TCSDs also allow the specification of timing constraints on events. To this end, the diagram type supports partition lines $\tau \in \Tcal$ with $\tau = (e_{1}, \dots, e_{m}, \delta)$ which are annotated with the timing information $\delta$. A partition line cuts through all instance lines by introducing a new event in each line. 
Each TCSD has an implicit partition line $\tau_0$ with $\delta = 0$ which indicates the start of the diagram. Every  time stamp  of following partition lines is relative to this initial line. 
The semantics are that every event before a partition line event has to happen before the annotated time stamp $\delta$ and every following event at least after $\delta$ time. 

The consistency of partition lines is ensured by additional properties. The uniqueness property (1) requires that there are no two partition lines with the same time stamp. The completeness property (2) ensures that all instance lines have a partition line event separating their own events. The ordering property (3) ensures that the annotated times on all partition lines are in an ascending order. The last property (4) prevents the intersection of partition lines with fragments.

For the ordering of two partition lines we will write $\tau_1 < \tau_2$ as a short form for comparing the time steps $\delta_1 < \delta_2$.

To express timing constraints within a (sub-)fragment of the SUT instance line, a TCSD supports the concept of timeouts, which are represented by the set $\Ccal$. Each timeout is a tuple $c=(e_1, e_2, \delta)$ consisting of two ordered events between which at most $\delta$ time units may pass. A timeout may be used within a subfragment, but both events of it must be part of the same fragment operand.

\begin{thm:def}[Test Case Sequence Diagram] 
\label{def:tcsd}
A test case sequence diagram (TCSD) is a tuple $TC = (d,I,E,<,\Sigma_{msg},M,F,X,Exp, sut, \Tcal, \Ccal)$ where:
\begin{itemize}
	\item $(d,I,E,<,\Sigma_{msg},M,F,X,Exp)$ is a sequence diagram;
	\item $sut \in I$ is the system under test instance line;
	\item for every $m\in M$ with $m=(e_1, l,e_2): (e_1 \in E_{sut} \wedge e_2 \notin E_{sut}) \vee (e_2 \in E_{sut} \wedge e_1 \notin E_{sut})$
	\item $\Tcal \subseteq E^{|I|} \times \N_0 $ is the set of time partition lines, \st for every $\tau_1, \tau_2 \in \Tcal$ with $\tau_1 = (e_{11},\dots, e_{1m},\delta_1)$ and $\tau_2 = (e_{21},\dots, e_{2m},\delta_2)$:
	\begin{enumerate}
	\item uniqueness: $\delta_1 = \delta_2 \implies \tau_1 = \tau_2$
	\item completeness: $\forall e_{1i}, e_{1j} :  i\neq j \Leftrightarrow E_i \neq E_j$
	\item ordering: $\forall j : (\delta_1 < \delta_2) \Leftrightarrow (e_{1j}, e_{2j}) \in E_j \implies (e_{1j}, e_{2j}) \in <_j;$
	\item no fragment cutting: $\forall f_1 \in$ with $op(f_1) = (o_1,n_1), \forall n < n_1: e_{11},\dots, e_{1m} \notin ev(f_1,n)$
	\end{enumerate}
  \item $\Ccal$ is a set of timeouts $\Ccal \subseteq E_{sut} \times E_{sut} \times \N$, \st for every $c=(e_1, e_2, \delta)$ with $e_1, e_2 \in E_{sut}$:
  \begin{enumerate}
	\item ordered: $e_1 < e_2$
	\item same fragment: $\forall f \in F $ with $op(f)=(o_1, n_1): e_1\in ev(f,x) \Leftrightarrow e_2\in ev(f,x)$ for any $x\leq n_1$
\end{enumerate}
\end{itemize}  
\end{thm:def}

\subsection{Timed-arc Petri Nets}
\label{ssec:pn}
In this section we will recall the formal definition of TAPN  and introduce our notation which will be used in the rest of the document.

\begin{thm:def}[Timed-arc Petri nets (based-on \cite{byg2009b}]
A \emph{timed-arc Petri net with transport arcs} (TAPN) is a tuple $N=(P,T,R,c, R_{tarc},c_{tarc}, \iota)$, where:
\begin{itemize}
	\item $P$ is a finite set of places;
	\item $T$ is a finite set of transitions $(P \cap T = \emptyset)$;
	\item $R$ is the flow relation $(R \subseteq (P \times T) \cup (T \times P))$;
	\item $c$ is a function that associates a time interval to each arc $(p,t)$ in $R$, \st $c : R\vert_{P\times T} \rightarrow \Ical$;
	\item $R_{tarc} \subseteq (P \times T \times P)$ is the set of \emph{transportation arcs} that satisfy for all $(p,t,p')\in R_{tarc}$ and all $r \in P$: 
$((p, t, r)\in R_{tarc} \implies p'=r) \wedge ((r, t, p')\in R_{tarc} \implies p=r) \wedge (p,t)\notin R \wedge (t,p')\notin R$ 
	\item $c_{tarc} : R_{tarc} \rightarrow \Ical$ associates a time interval to every transportation arc;
	\item $\iota: P \rightarrow \Ical_{Inv}$ assigns time intervals as invariants to places.
\end{itemize} 
 
\end{thm:def}

\noindent
The TAPN can be seen as a directed bipartite graph of separated places $P$ and transitions $T$. The interconnection of the net is given by two flow relations. $R$ defines arcs between places and transitions in both ways as known from conventional Petri nets, whereas $R_{tarc}$ defines so-called transportation arcs. The main structural difference between those two kinds of arcs is that transportation arcs are always triples from a place over a transition to a place. The other arcs are separate tuple for arcs to a transition $(p,t)$ or arcs from a transition $(t,p)$.
The additional condition on transportation arcs imposes, that there is at most one transportation arc between any two places.

In contrast to the original definition \cite{byg2009b}, we limit the supported time intervals $\Ical$ and $\Ical_{Inv}$ to be only closed $[\cdot,\cdot]$ or right-open $[\cdot,\cdot)$ intervals over $\N_0 \times (\N_0 \cup \lbrace \infty \rbrace)$. Other kinds of intervals are not relevant for our virtual integration analysis. In addition, invariants are also not considered and we assume the default invariant $[0, \infty)$ for every place.

In the following we will use $p \pnarc{[\timeoutbegin, \timeoutend)} t$ as a short notation for the arc $(p,t)$ with an associated time interval $[\timeoutbegin, \timeoutend )$ and $p \pnarc{[\timeoutbegin, \timeoutend)} t \emptyarc p'$ for a sequence in a TAPN $N$ where $(p,t),(t,p') \in R_N$ and $c_N(p,t) = [\timeoutbegin, \timeoutend)$. For transportation arcs we use the notation $p \tarc{[\timeoutbegin, \timeoutend)} t \emptytarc ~p'$ respectively.

The state of a Petri net is defined by its current placed tokens (marking). In a TAPN each token has an individual age which increases over time.
Formally a marking $M$ on a TAPN $N$ is a function $M: P \rightarrow 2^{\Rposo}$ which assigns every place $p\in P$ a set of positive real numbered tokens, \st each token $x\in M(p)$ of a marking fulfills the invariant of its assigned place: $x \in \iota(p)$. Only markings with a finite number of tokens are considered in this paper.

A marked TAPN is a tuple $(N,M_0)$ of a TAPN $N$ and a marking $M_0$ over the places of $N$. The dynamics are defined over changes of this marking, which follow the flow relations $R$ and $R_{tarc}$. A transition $t$ is said to be \emph{enabled} if there exists at least one token in each of the places connected to its incoming arcs according to $R$ for normal arcs and $R_{tarc}$ in case of transportation arcs and these tokens are in the corresponding timing interval of $c$ and $c_{tarc}$ respectively. 
In addition, the tokens of normal arcs and transportation arcs have to fulfill the invariants of the target places. An enabled transition can \emph{fire} by consuming exactly one token of matching age from each of its incoming arcs and producing one new token of each of its outgoing arcs. The age of these newly created token is either $0$, if it is produced by a normal arc, or the age of the consumed token, if it is produced by a transportation arc.
Instead of firing a transition, a TAPN can perform a so called time delay in which the age of all token of the current marking is increased by the same timespan.
The time delay is valid if all token still fulfill the invariants of their corresponding places.
We write $M \pnarc{t} M'$ if the marking $M'$ is reached from $M$ by consuming $t$ time units and firing enabled transitions of the TAPN.
A marking $M_k$ is said to be reachable from $M_0$ within $k$ steps, if there is a sequence $M_i \pnarc{t_i} M_{i+1}$ for $0\leq i \leq k$ and $i,k \in \N_0$.

For the construction of TAPNs from sequence diagrams, we need to be able to identify transitions with messages of the diagram. Therefore, we define a labeling function $\lambda_{trans}: T \rightarrow \Sigma_{trans}$, which associates each transition of a TAPN with a label from a labeling alphabet $\Sigma_{trans}$.

%% file: translation.tex
\section{Virtual Integration of Sequence Diagrams}
\label{sec:virtualIntegration}

In the virtual integration scenario all TCSDs of the individual components are combined according to the interconnections of the ports provided by the system architecture. The procedure consists of three steps: First, the TCSDs are translated into TAPNs, which mimic the occurrences of the SUT events and impose the same timing constraints; Second, the individual TAPNs are combined to a single TAPN by synchronizing the communicated messages according to the architecture; And third, a consistency analysis is performed which checks if it is possible to successfully execute the combined TAPN.

\subsection{Translation of TCSDs to TAPNs}
\label{ssec:translate}
\begin{figure}[htb]
	\centering
		\includegraphics[width=0.8\textwidth]{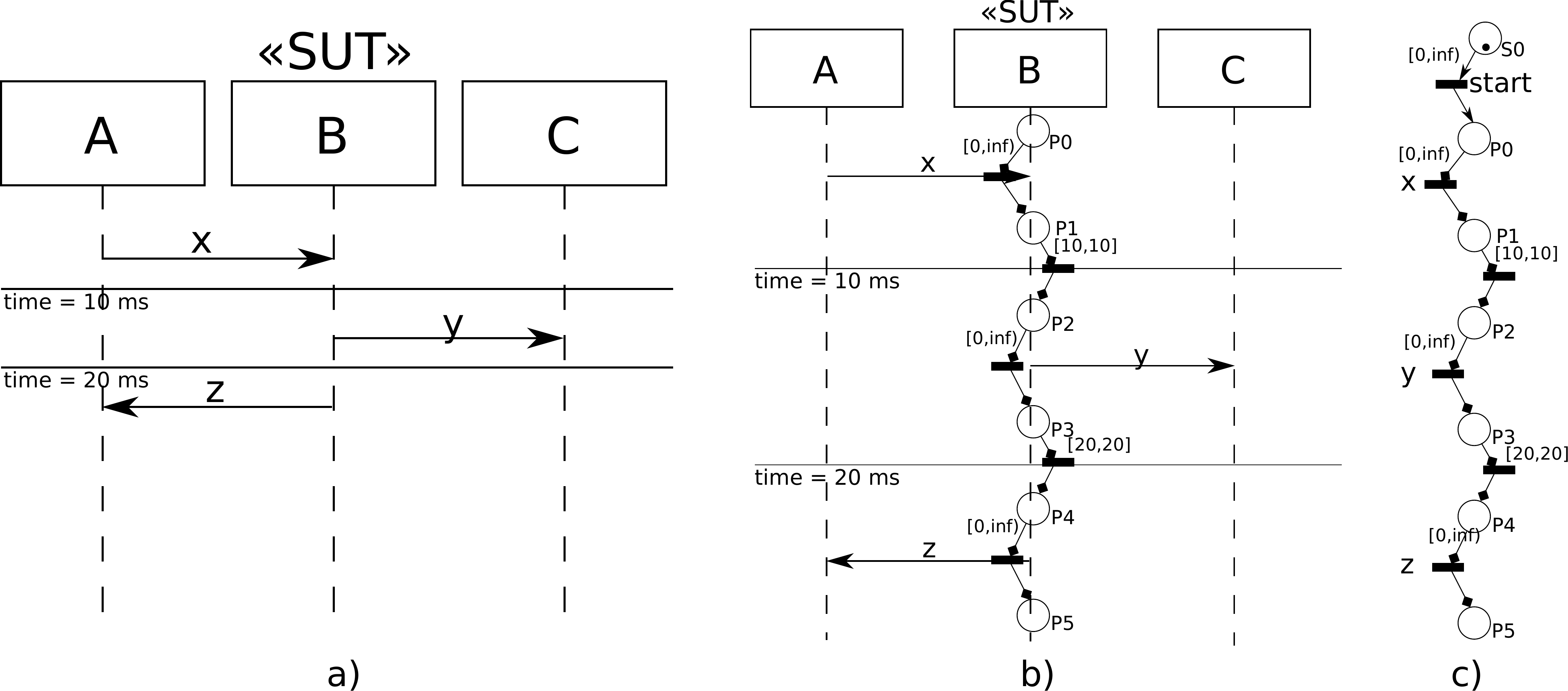}
	\caption{Example of the translation procedure of a TCSD to a TAPN. Subfigure \emph{a)} shows a basic TCSD with the sequence of messages \texttt{x, y, z} and required timing-constraints. Subfigure \emph{b)} shows how the message events relate to labeled transitions on the main transportation arc sequence of the  TAPN. The timing-constraints are enforced by the interval guards. Subfigure \emph{c)} shows the resulting marked TAPN including a place with a token and the delay transition at the beginning.
	}
	\label{fig:translationIdea}
\end{figure}

For the translation of TCSDs to TAPNs we define a set of translation rules. 
The general idea is to construct a sequence of transitions (main sequence) in which each transition represents exactly one event of the SUT instance line. In the TAPN a single token is transported among this sequence and mimics the execution of the sequence diagram. 
The age of the token on the main sequence is restricted by timing constraints on the transitions according to the constraints on the partition line.
Figure \ref{fig:translationIdea} shows the idea of the construction of the main sequence.

Messages are represented as labels on transitions. They don't add any additional semantic to the net, but are used for synchronization with other nets. 
Fragments and timeout events on the other hand, are translated by adding branches (one for each operand) to the main line. Each branch starts at the \texttt{enter} event transition and is merged again with the main line on the corresponding \texttt{exit} transition.

The formal construction is defined by a set of translation rules. Each rule applies to one kind of event. Their input on the one hand is the TCSD $TC$ which has to be translated and an event $e_c \in E_{sut}$ of the SUT instance line which marks the current position in the diagram. On the other hand the other input is the so far partially generated TAPN $PN$ and a place $p_c$ to which the new Petri net elements have to be appended. The place $p_c$ is either the last place of the generated main sequence or the end of the current branch, if $e_c$ is part of a fragment operand.
The output of the rule is an extended net $PN'$ and the next event $e'_c$.

For the notation of the rules we will mark every new TAPN element added to $PN$ with a prime. We will also use the function $next$ to indicate the immediate following event within the SUT instance line according to the $<$ relation of the diagram. In case there is just a partial ordering because of multiple operands in a fragment, the rules will iterate the events of the operands sequentially. Intuitively, the transformation rule iterates over all events on the SUT instance line according to the graphical notation.
If $next(e) = \emptyset$ then the event $e$ is the last event on the instance line. In addition, the functions $first$ and $last$ are used to identify the first respectively the last event of a operand according to the ordering relation.
Each rule consists of a formal description and a figure of the TAPN artifacts created by the rule. Places in \emph{red} mark the attachment point to the prior constructed TAPN elements. Dashed lines indicate segments which have to be further extended.

The transformation process consists of a succeeding application of the different translation rules and an iterative construction of the corresponding TAPN. 
Initially, Translation Rule \ref{rule:init} is applied which creates the initial places and the marking $M_0$.
Afterwards the Translation Rules \ref{rule:msg}--\ref{rule:timeoutend} are applied according to the type of the current event $e_c$. They extend $PN$ with a TAPN fragment modeling the sequence diagram event type until the last event has been handled ($next(e_c)=\emptyset$).
Then the final Translation Rule \ref{rule:termination} can be applied, which creates the marking $M_{target}$ indicating the target state to be reached from the TAPN $(PN,M_0)$.

\vspace{1em}
\noindent
\begin{minipage}[c]{0.68\textwidth}
\begin{thm:rule}[Initial]
\label{rule:init}
Let $e_c \in E_{sut}$ and $PN$ be the empty TAPN.
Create $p_c \InfArrow t' \emptyarc p'_c$ with $\lambda'(t') = \epsilon$, $e'_c = e_c$, and an initial marking $M_0$ consisting of a single token on place $p_c$ with age $0$.
\end{thm:rule}
\end{minipage}
\begin{minipage}[c]{0.3\textwidth}
\begin{center}
\fbox{
\begin{tikzpicture}[node distance=1.5cm,>=stealth,bend angle=45,auto]
\begin{scope}
\node [place,tokens=1]		(pc)     	[label=below:$p_c$]									{};

\node [transition] 				(tc2)			[right of=pc, label=below:$t'$]		{}
	edge [pre] 	node 		  [above]{$[0,\infty)$} (pc)  				;
	
\node [place] 						(pc2)    	[right of=tc2, label=below:$p'_c$]	{}
	edge [pre] 		(tc2);
\end{scope}	
\end{tikzpicture}
}
\end{center}
\end{minipage}

\vspace{1em}
\noindent
Multiple sequence diagrams may be executed with an arbitrary time offset. Therefore, we prefix the main sequence with a transition, such that the TAPN can initially wait. 
The Translation Rule \ref{rule:init} adds a transition with normal arcs and no timing constraint before the first event of the TCSD.
This models an arbitrary offset between the starting times of all TCSDs, \eg Figure \ref{fig:translationIdea} \emph{c)} $S0 \InfArrow start \emptyarc P0$.

\vspace{1em}
\noindent
\begin{minipage}[c]{0.68\textwidth}
\begin{thm:rule}[Send/Receive Message Event]
\label{rule:msg}
Let $e_c \in E_{sut}$ where $\exists m\in M$, \st $m=(e_c,l,e)$ or $m=(e,l,e_c)$ and $p_c \in P$. 
Append $p_c \InfTarc t' \emptytarc p'_c$ with $\lambda'(t') = l$ and $e'_c = next(e_c)$
\end{thm:rule}
\end{minipage}
\begin{minipage}[c]{0.3\textwidth}
\begin{center}
\fbox{
\begin{tikzpicture}[node distance=1.5cm,>=stealth,bend angle=45,auto]
\begin{scope}
\node [red place] 						(pc)     	[label=below:$p_c$]									{};

\node [transition] 				(tc2)			[right of=pc, label=below:$t'$, label=above:$l$]		{}
	edge [pre, >=diamond] 	node 		  [above]{$[0,\infty)$} (pc)  				;
	
\node [place] 						(pc2)    	[right of=tc2, label=below:$p'_c$]	{}
	edge [pre,>=diamond] 		(tc2);
\end{scope}	
\end{tikzpicture}
}
\end{center}
\end{minipage}

\vspace{4em}
\noindent
After the application of the initial rule, the other rules generate transportation arcs according to the events on the instance line.
Translation Rule \ref{rule:msg} models the sending and receiving of messages by adding a transportation arc to the main sequence with no additional timing constraints.
The created transition $t'$ is labeled with the label of the message.
This label identifies the message (\eg it contains the port identifier and the sent/received content) and is later used for synchronization with other TAPNs, \eg Figure \ref{fig:example-sd} \(m = (e_{11},l_1, e_{21})\).


\vspace{1em}
\noindent
\begin{minipage}[c]{0.68\textwidth}
\begin{thm:rule}[Partition Line Event]
\label{rule:partition}
Let $e_c \in E_{sut}$ where $\exists \tau = (e_1,\dots, e_n,e_c,e_m,\dots, e_l,\delta)$ and $p_c \in P$. 
Append $p_c \tarc{[\delta,\delta]} t'_c \emptytarc p_c'$ with $\lambda'(t') = \epsilon$ and $e'_c = next(e_c)$
\end{thm:rule}
\end{minipage}
\begin{minipage}[c]{0.3\textwidth}
\begin{center}
\fbox{
\begin{tikzpicture}[node distance=1.5cm,>=stealth,bend angle=45,auto]
\begin{scope}
\node [red place] 						(pc)     	[label=below:$p_c$]									{};

\node [transition] 				(tc2)			[right of=pc, label=below:$t'_c$]		{}
	edge [pre, >=diamond] 	node 		  [above]{$[\delta,\delta]$} (pc)  				;
	
\node [place] 						(pc2)    	[right of=tc2, label=below:$p'_c$]	{}
	edge [pre,>=diamond] 		(tc2);
\end{scope}	
\end{tikzpicture}
}
\end{center}
\end{minipage}

\vspace{1em}
\noindent
The timing constraints imposed by partition lines are also encoded by adding a single transportation arc to the main sequence of the graph. The interval $[\delta, \delta]$ limits the age of the main token to exactly $\delta$ in order to fire this transition. This represents the semantics of partition lines, \st all events before the partition line have to be executed before time stamp $\delta$ and all events after the partition line can only be executed after time stamp $\delta$.

\vspace{1em}
\noindent
\begin{minipage}[c]{0.68\textwidth}
\begin{thm:rule}[Par-/Alt-/Opt-Fragment Event]
\label{rule:par}
Let $e_c \in E_{sut}$ where $\exists f\in F$ with $op(f)=(\Omega, N)$ with $\Omega \in \lbrace \texttt{par,alt,opt}\rbrace$ and $e_c$ is \texttt{enter} event of $f$:\\
Append $p_c \InfTarc t'_{f-start} \emptytarc p'_{f-start} \InfTarc t'_{f-end} \InfTarc p'_{f-end}$.\\
For each operand id $n<N$ do:
\begin{enumerate}
	\item Append $t'_{f-start} \emptyarc p'_{op-start}$ 
	\item Apply Rules \ref{rule:msg}--\ref{rule:timeoutend} starting with $e_c=first(ev(f,n))$ and $p_c=p'_{op-start}$ until $e'_c \notin ev(f,n)$
	\item Append $p'_{op-end}\InfArrow t'_{f-end}$ with $p'_{op-end}=p'_c$ after step 2.
\end{enumerate}

\noindent
Finally, set $p'_c=p'_{f-end}$ and $e'_c = next(e_c)$.\\
\end{thm:rule}
\end{minipage}
\begin{minipage}[c]{0.3\textwidth}
\centering
\fbox{
\includegraphics[width=0.82\textwidth]{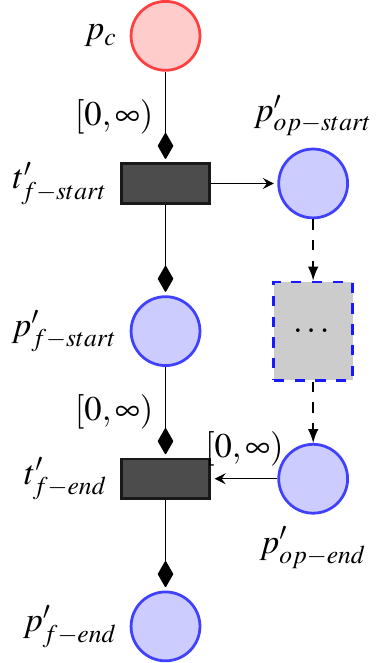}
}
%
%
%
%
%
%
%
%
%
\end{minipage}

\vspace{1em}
\noindent
Fragment blocks \texttt{alt, opt, par} \cite{uml2} and the possible hierarchical structures thereof are encoded by adding additional paths, branching from the main sequence for each operand of the fragment. The main sequence is extended by two transitions. The first transition $t'_{f-start}$ represents the \texttt{enter} event of the fragment and the second $t'_{f-end}$ the corresponding \texttt{exit} event. 
A new branch starting from $t'_{f-start}$ and ending at $t'_{f-end}$ is added to the main sequence for each of the operands of the fragment. The branches itself are constructed according to the normal Translation Rules \ref{rule:msg}--\ref{rule:timeoutend}.
This construction effectively reduces the \texttt{alt} and the \texttt{opt} fragments to the parallel execution construction of \texttt{par}. The motivation is that the resulting TAPN shall represent all possible execution paths of the original TCSD in order to provide all possible synchronization points for the virtual integration with other TAPN. 

\vspace{1em}
\noindent
\begin{minipage}[c]{0.68\textwidth}
\begin{thm:rule}[Strict-Fragment Event]
\label{rule:strict}
Let $e_c \in E_{sut}$ where $\exists f\in F$, $op(f)=(\Omega, N)$ with $\Omega=\texttt{strict}$ and $e_c$ is
\texttt{enter} event of $f$: Set $e'_c=next(e_c)$
\end{thm:rule}
\end{minipage}

\vspace{1em}
\noindent
Translation Rule \ref{rule:strict} handles \texttt{strict} fragments \cite{uml2}, which is the default semantic in our construction. Therefore, it does not require any additional handling and the corresponding \texttt{enter} and \texttt{exit} events can be skipped.

\vspace{1em}
\noindent
\begin{minipage}[c]{0.68\textwidth}
\begin{thm:rule}[Loop-Fragment Event]
\label{rule:loop}
Let $e_c \in E_{sut}$ where $\exists f\in F$, $op(f)=(\Omega, 1)$ with $\Omega=\texttt{loop}$, $guard(f)=N$ with $N\in \N$, and $e_c$ is \texttt{enter} event of $f$:\\
Append $p_c \InfTarc t'_{f-start} \emptytarc p'_{f-start} \InfTarc t'_{f-end} \InfTarc p'_{f-end}$.

\begin{enumerate}
	\item Append $t'_{f-start} \emptyarc p'_{l-start}$ 
	\item Repeat $N$-times:
	\begin{itemize}
	\item Apply Rules \ref{rule:msg}--\ref{rule:timeoutend} starting with $e_c=first(ev(f,1))$ and $p_c=p'_{l-start}$ until $e'_c \notin ev(f,n)$
\end{itemize}
 	\item Append $p'_{l-end}\InfArrow t'_{f-end}$ with $p'_{l-end}=p'_c$ after step 2.
\end{enumerate}
Finally, set $p'_c=p'_{f-end}$ and $e'_c = next(e_c)$.\\
\end{thm:rule}
\end{minipage}
\begin{minipage}[c]{0.3\textwidth}
\centering
\fbox{
\includegraphics[width=0.82\textwidth]{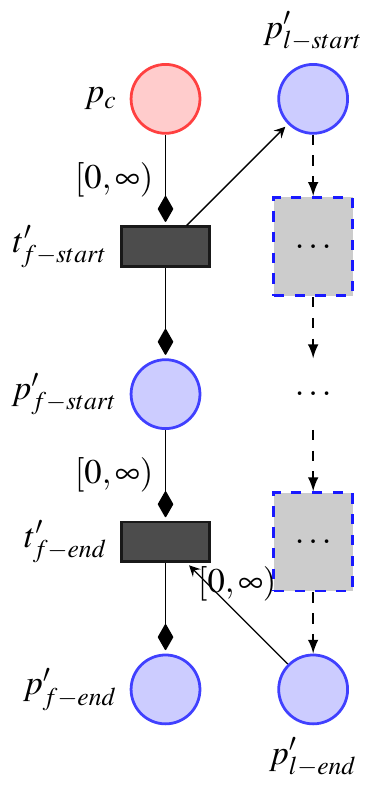}
}
%
%
%
%
%
%
%
%
%
%
\end{minipage}

\vspace{1em}
\noindent
The \texttt{loop} fragment is only supported if its guarding expression is a constant number. The construction described in Rule \ref{rule:loop} is similar to the construction used for the \texttt{par} fragments. Except it is limited to a single operand and instead of adding multiple branches, it extends the looping branch starting with $p'_{l-start}$ $N$-times where $N$ is the number of unrollings of the loop. 

\vspace{1em}
\noindent
\begin{minipage}[c]{0.68\textwidth}
\begin{thm:rule}[Timeout Event]
\label{rule:timeoutend}
Let $e_c \in E_{sut}$ where $\exists c=(e_{start},e_{end},\delta) \in \Ccal$ with $e_c=e_{start}$:
\begin{enumerate}
	\item Append $p_c \InfTarc t'_{to-start} \emptytarc p'_{to-start}$
	\item Apply Rules \ref{rule:msg}--\ref{rule:timeoutend} starting with $e_c=next(e_c)$ and $p_c=p'_{to-start}$ until $e'_c = e_{end}$
	
	\item Append $p'_{to-end} \InfTarc t'_{to-end} \emptytarc p'_c$ with $p'_{to-end}=p'_c$ after step 2.
	\item Append $t'_{to-start} \emptyarc p_{wait} \pnarc{[0,\delta]} t'_{to-end}$

\end{enumerate}
\end{thm:rule}
\end{minipage}
\begin{minipage}[c]{0.3\textwidth}
\centering
\fbox{
\includegraphics[width=0.82\textwidth]{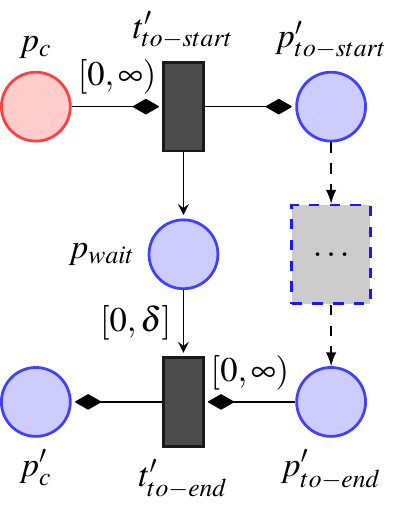}
}
%
%
%
%
%
%
%
%
%
%
%
%
\end{minipage}

\vspace{1em}
\noindent
Timeout events define a timing constraint between the occurrences of the start event $e_{start}$ and the end event $e_{end}$, \eg the timeout operator \((e_{25}, e_{29}, 5ms)\) in Figure \ref{fig:example-sd} constraints the time between the occurences of the events \(e_{26}, e_{27}, e_{28}\) to \(5ms\).
The construction described in Translation Rule \ref{rule:timeoutend} creates a new transportation arc on the main sequence for each of these events and a connected place with the corresponding timing constraint as interval.
In contrast to fragment event rules, this rules constructs all events between $e_{start}$ and $e_{end}$ directly on the main sequence rather than on a branch.

\vspace{1em}
\noindent
\begin{minipage}[c]{0.68\textwidth}
\begin{thm:rule}[Termination]
\label{rule:termination}
Let $e_c \in E_{sut}$ of $SD$ and $p_c \in P$ of $PN$. 
If $next(e_c) = \emptyset$ then $PN$ is the transformed TAPN of $SD$ with correspondence in timed reachability of the marked TAPN $(PN,M_0)$ to a marking $M_{target}$, where $M_0$ is the initial constructed marking and $M_{target}$ is a marking consisting of a single token (with arbitrary age) on $p_c$.
\end{thm:rule}
\end{minipage}

\vspace{1em}
\noindent
The translation process terminates with the application of Translation Rule \ref{rule:termination}. This rule is only applicable if the end of the SUT instance line is reached ($next(e_c)=\emptyset$) and it creates the target Marking $M_{target}$ representing the state after the execution of the TCSD in the constructed TAPN.

\subsection{Synchronization of TAPN}

The virtual integration is done according to a system architecture $A$, which instantiates components and specifies how they are interconnected. 
Each component is associated with its own set of test cases, specified as TCSDs. 
Within a TCSD the component itself is identified as the SUT instance line. 
The other instance lines are test components, which represent virtual communication partners for the ports of the component. 

Given the connection between the ports of the components within the architecture it is possible to create a mapping of all test components to existing components of the architecture.
This mapping directly corresponds to a mapping of SUT instance lines and test component instance lines.

We write $i_1 \cong_A i_2$ for two instance lines $i_1 \in I_1$ and $i_2 \in I_2$ to indicate that $i_1$ and $i_2$ are in a mapping relation according to the interconnections of the architecture $A$. 

The synchronization of these instance lines is done on the basis of the messages transmitted on the common ports. For the compatibility of messages we define a similar relation in Definition \ref{def:compatible}. Two messages are compatible if their source and target instance lines are in the mapping relation $\cong_A$.

\begin{thm:def}[Compatibility of Messages]
\label{def:compatible}
Let $TC_1$, $TC_2$ be two TCSD with $TC_1 \neq TC_2$ and $A$ be an architecture connecting the components:\\
The compatibility $\cong_A$ of two messages $m_1 = (e_{11},l_1, e_{12}) \in M_{TC_1}$ 
with $e_{11} \in E_{i_{11}}, e_{12} \in E_{i_{12}} $ 
and $m_2 = (e_{21},l_2, e_{22}) \in M_{TC_2}$ 
with $e_{21} \in E_{i_{21}}, e_{22} \in E_{i_{22}} $ 
is defined as: $(m_1 \cong_A m_2) := (i_{11} \cong_A i_{21}) \wedge (l_1 = l_2) \wedge (i_{12} \cong_A i_{22})$
\end{thm:def}

\begin{figure}[h!]%
\centering
\fbox{
\includegraphics[width=0.8\textwidth]{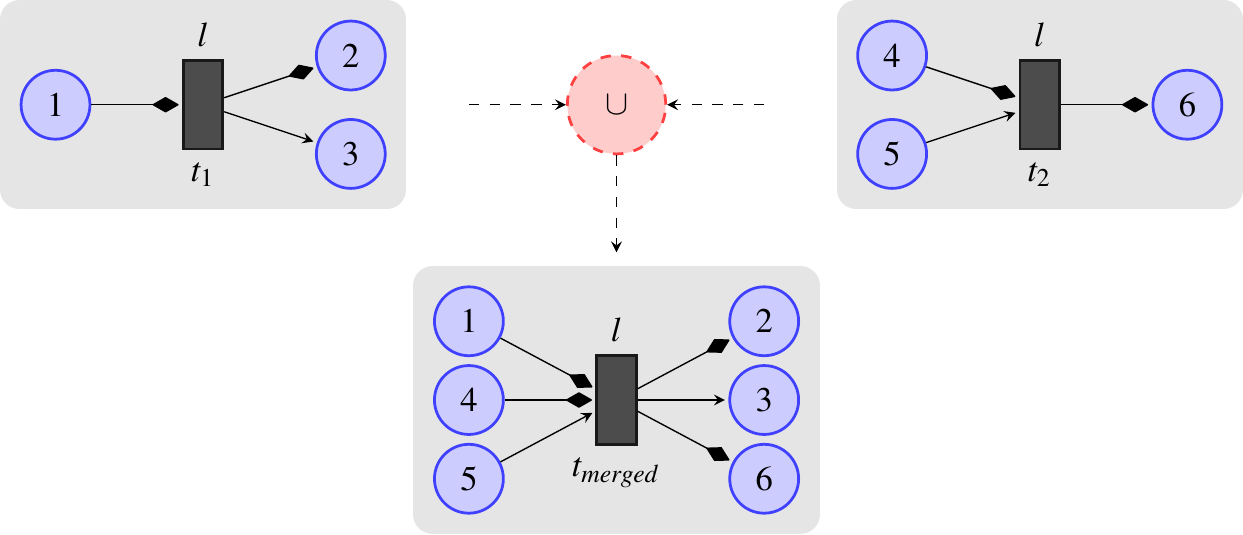}
}

\caption{Example of the combination of two (partial) TAPNs. The two transitions $t_1$ and $t_2$ of the nets on the top are synchronized by replacing the transitions with a newly created transition $t_{merged}$. All arcs (transportation and normal) to $t_1$ or $t_2$ are redirected to $t_{merged}$. The places in this figure are numbered to identify them in the merged TAPN.}%
\label{fig:merging}%
\end{figure}

\noindent
Given this definition of compatibility, we can combine multiple TCSDs by synchronizing all compatible messages of the two diagrams. This synchronization process is in general not unique \eg if there are multiple instances of messages with the same label). In this case all combinations of potential synchronization points have to be considered. 
To this end, we transform each individual TCSD into its corresponding TAPN. In the TAPN representation, each message is represented by a unique transition (see Translation Rule \ref{rule:msg}). In case of a synchronization, the two transitions representing the compatible messages can be combined to a single new transition as depicted in Figure \ref{fig:merging}.
Formally, all sets and functions of the two TAPN are merged, new transitions are introduced replacing the individual synchronized transitions, and the arcs are redirected to the new transitions.

\subsection{Consistency Analysis}
After all individual TAPNs are merged to a single net, the final step of the virtual integration analysis is to determine if this TAPN is consistent.
Consistency in this case means that the target marking of the combined TAPN (the union of all individual target markings) is reachable from its initial marking (the union of all initial markings).

This procedure detects timing constraint violations and ordering inconsistencies of messages. 
Timing constraints in the constructed TAPN relate to interval bounds (lower and upper) on the age of the token on the individual main sequences and the token of timeout operator branches (see Translation Rule \ref{rule:timeoutend}). 
If two or more nets have to synchronize on a common transition the ages of these token must still fulfill their guards even if the token have to wait additional time for the synchronization. 
Formally, the analysis has to check if all interval constraints on synchronized transitions have a non-empty intersection. This ensures that at least one successful execution of the integrated components fulfills all timing constraints. 

The second detectable kind of inconsistency is the problem regarding the ordering of messages. 
If for example one test case defines a strict ordering of two messages $m_1$ and $m_2$ and a second test case specifies the opposite ordering of first $m_2$ and then $m_1$, the merged TAPN can never reach its target marking. 
The messages $m_1/m_2$ in both test cases are translated into transitions $t_1/t_2$ according to Transition Rule \ref{rule:msg}. 
After synchronizing the transitions, the net has a classical deadlock in its transitions: $t_1$ can by construction only fire after $t_2$ fired, but $t_2$ has to wait until $t_1$ fired.

In case there are multiple possibilities to synchronize two nets, the analysis has to consider all possible points of synchronization. For our analysis we consider it as sufficient, if at least one of the combinations succeeds the reachability analysis. Alternatively, one could impose a stronger consistency concept if it is required that all possible combinations pass the analysis.

In general test cases define only the fragment of the total component behavior which is relevant for the test case. Therefore, the specified message sequences are in most cases incomplete. This can lead to a \emph{false} inconsistent result of the analysis. The prior mentioned inconsistent sequence $m_1$ then $m_2$ may be consistent if we assume that the second test case just neglected a first occurrence of $m_1$ \eg the full sequence would be $m_1,m_2,m_1$. Therefore, inconsistency results may just be cases of underspecified test cases.

%% file: evaluation.tex
\section{Evaluation}
\label{sec:evaluation}
\begin{figure}[h]
	\centering
	\subfloat[Architecture]{\label{fig:arch}\includegraphics[width=0.4\textwidth]{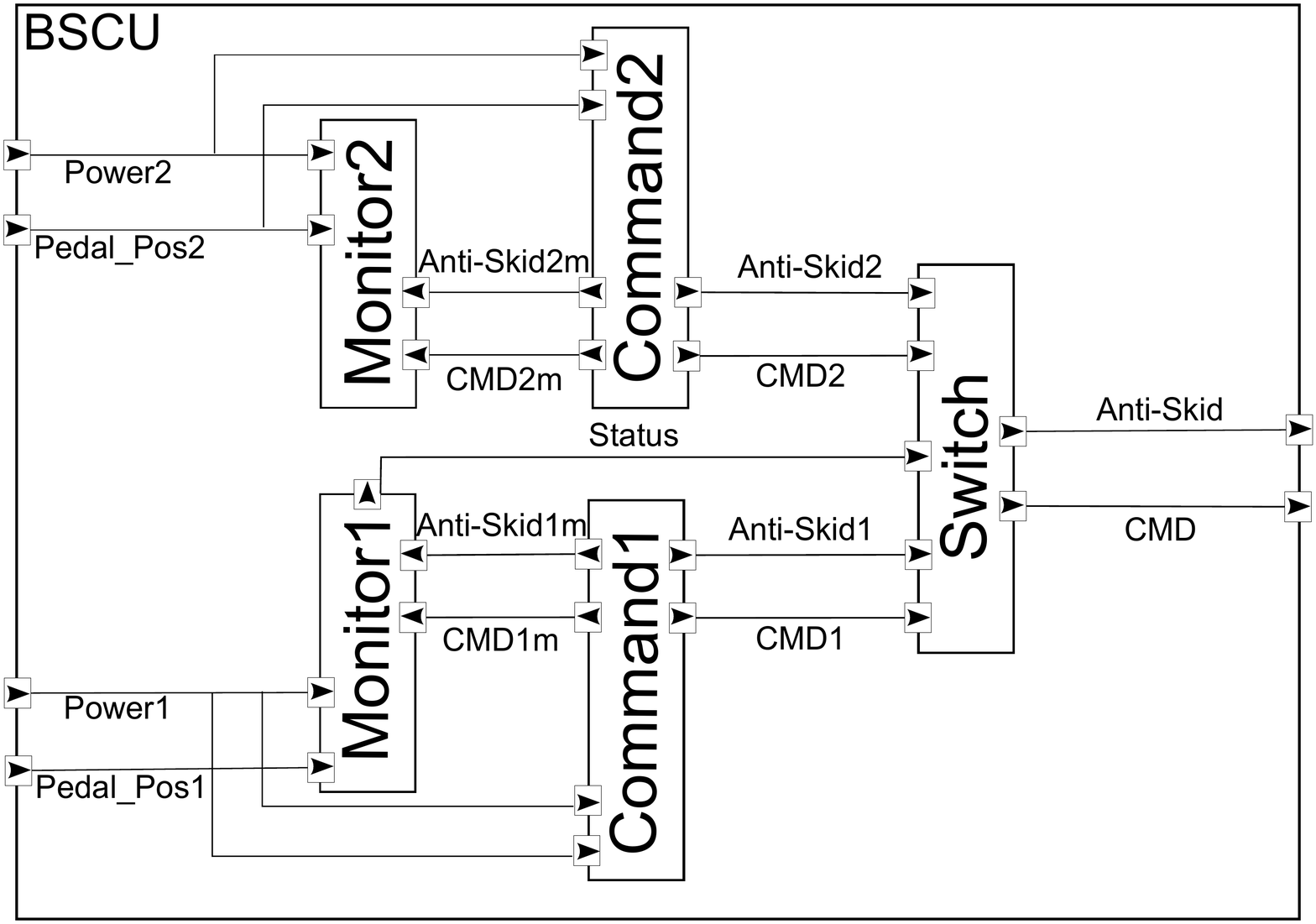}} \hspace{0.1\textwidth}
	\subfloat[Test case with SUT \texttt{Command1}]{\label{fig:seq1}\includegraphics[width=0.4\textwidth]{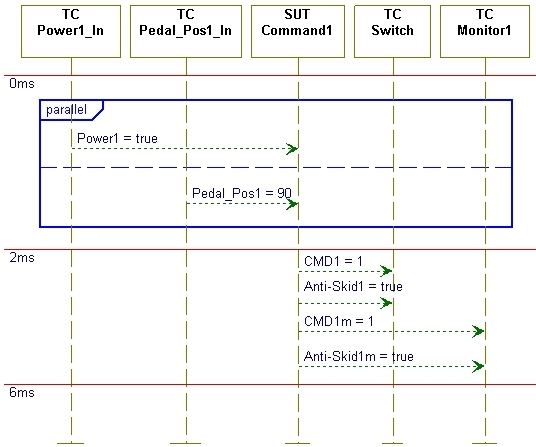}}

	\subfloat[Test case with SUT \texttt{Monitor1}]{\label{fig:seq2}\includegraphics[width=0.4\textwidth]{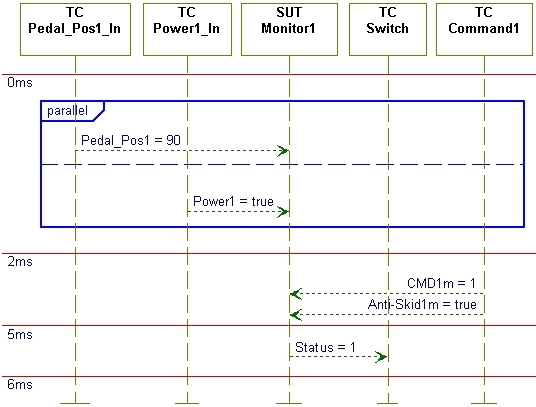}} \hspace{0.1\textwidth}
	\subfloat[Test case with SUT \texttt{Switch}]{\label{fig:seq3}\includegraphics[width=0.4\textwidth]{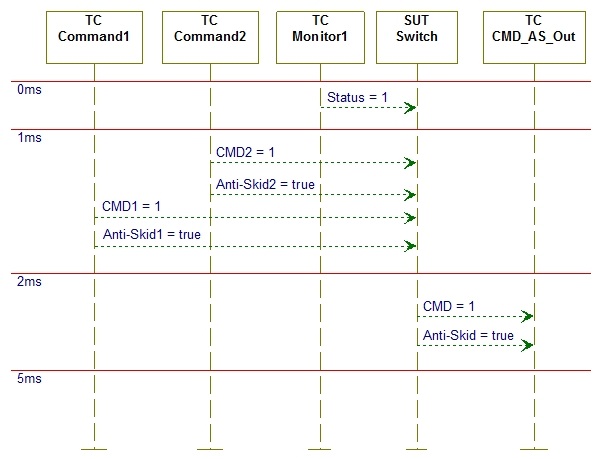}}
	\caption{Subfigure a) shows the component model of the BSCU. Subfigures b) - d) depict test cases used for the demonstration of our approach. The SUTs and test components of each test case are mapped on the corresponding component in the shown architecture.} 
	\label{fig:set_of_tc}
\end{figure}

We demonstrate the virtual integration analysis presented in this paper on an example that is complex enough for our purpose but still has an acceptable degree of simplicity to understand the context.
The Brake System Control Unit (BSCU) is part of a Wheel Braking System (WBS) example that was used to describe the safety assessment for certification of civil aircraft in the SAE standard ARP4761 \cite{ARP47611996}.
The architecture of the BSCU, which consists of two redundant subsystems, is shown in Figure \ref{fig:arch}.

\begin{figure}[h]
\centering
\includegraphics[width=0.78\textwidth]{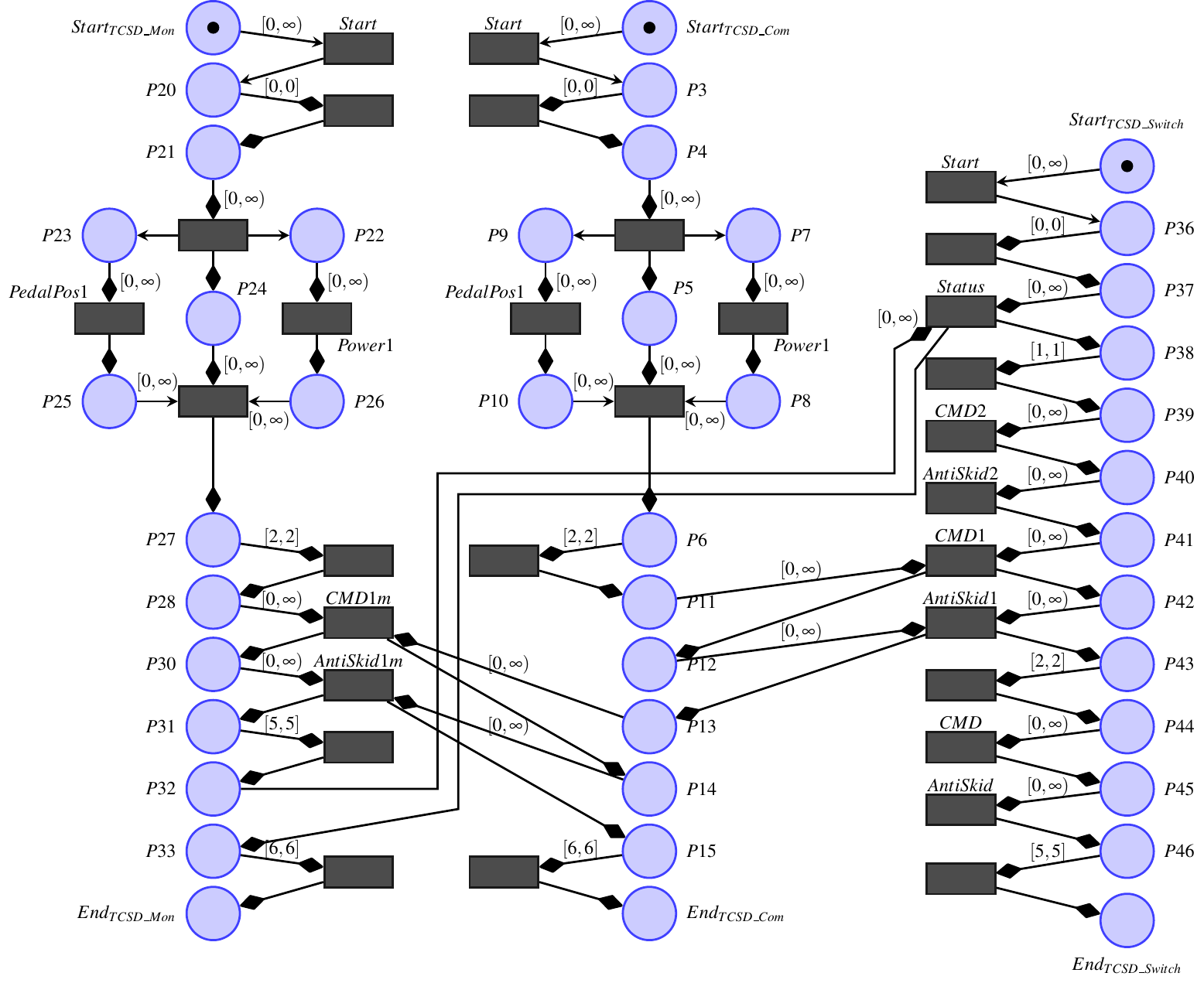}
\vspace{-0.5cm}
\caption{Merged timed-arc Petri net of the test case sequence diagrams.}
\label{fig:tapn}
\end{figure}

We created a set of test cases to demonstrate our approach.
These test cases represent the interaction of specific components of the
BSCU, \eg a test case for the \textit{Command} unit (Figure  \ref{fig:seq1}),
\textit{Monitor} (Figure \ref{fig:seq2}), and the \textit{Switch} (Figure
\ref{fig:seq3}).
Applying our approach, each of these sequence diagrams is translated into one
timed-arc Petri net.
The three resulting Petri nets are then being merged into one single TAPN,
see Figure \ref{fig:tapn}.

The translation rules introduced in section \ref{ssec:translate} can
be identified in Figure \ref{fig:tapn}.
For example a translation of a partition line (Translation Rule
\ref{rule:partition}) can be seen in the top left corner ($P20 \tarc{[0,0]}
t \emptytarc P21$).
Another example is a realisation of the \textit{par}-operator (Translation Rule
\ref{rule:par}).
Its main sequence is represented by $P21 \InfTarc t_1 \emptytarc P24 \InfTarc t_2 \emptytarc
P27$ and its branches are starting at the places $P22$ and \(P23\).

Performing the consistency analysis means that the target marking M =
\((End_{TCSD\_Mon},\)\(End_{TCSD_Com},\) \(End_{TCSD_Switch})\) is reachable.
In this case, the target marking is not reachable because of a deadlock,
caused by the transitions \textit{Status}, \textit{CMD1m},
\textit{AntiSkid1m}, \textit{CMD1} and \textit{AntiSkid1}.
Since this is detected by our analysis, we are able to fix this error
before the integration of the real system.

%% file: conclusion.tex
\section{Conclusion}
\label{sec:conclusion}

We presented an approach to analyze \sd -based test cases. Therefore, a concept of
test case \sd s was introduced, which allows to annotate timing information to test cases.
In order to formalize these test cases we extended the
\sd{} formalism of Bowles by the additional timing information.
In addition, we adapted the Petri net formalism of Byg to
represent the needed test case elements. For the translation of a TCSD
into a TAPN, a set of translation rules was presented. These TAPNs can be merged
into one single Petri net. On the basis of the merged
Petri net, we were able to analyse if a set of test cases is consistent in
the sense of ordering and timing behavior. 

The applicability of the approach was demonstrated using a example from the
ARP 4761. We used IBM Rational Rhapsody to specify test cases for this BSCU and
developed a prototype to extract the \sd -based specifications.
It also translates them into a TAPN to enable the analysis.

In the future we want to evaluate this approach on a larger scale design process, in
which our TCSDs are used for requirements specification as well as for test cases.
In addition, we want to extend the scope of the analysis to enable support
for life sequence charts \cite{Damm2011}. This will enable the integration
of the analysis into the early stages of the development process, \eg to
analyse requirements for early verification.
Other forms of synchronization of TAPNs, \eg check for time
interval inclusion in request/response scenarios are also planned.

%% file: petrinets.bbl
\begin{thebibliography}{10}
\providecommand{\bibitemdeclare}[2]{}
\providecommand{\surnamestart}{}
\providecommand{\surnameend}{}
\providecommand{\urlprefix}{Available at }
\providecommand{\url}[1]{\texttt{#1}}
\providecommand{\href}[2]{\texttt{#2}}
\providecommand{\urlalt}[2]{\href{#1}{#2}}
\providecommand{\doi}[1]{doi:\urlalt{http://dx.doi.org/#1}{#1}}
\providecommand{\bibinfo}[2]{#2}

\bibitemdeclare{article}{ARP47611996}
\bibitem{ARP47611996}
\bibinfo{author}{SAE \surnamestart ARP4761\surnameend} (\bibinfo{year}{1996}):
  \emph{\bibinfo{title}{{Guidelines and methods for conducting the safety
  assessment process on civil airborne systems and equipment}}}.
\newblock {\sl \bibinfo{journal}{SAE International}}, pp.
  \bibinfo{pages}{1--331}.

\bibitemdeclare{article}{Bowles2010}
\bibitem{Bowles2010}
\bibinfo{author}{Juliana \surnamestart Bowles\surnameend} \&
  \bibinfo{author}{Dulani \surnamestart Meedeniya\surnameend}
  (\bibinfo{year}{2010}): \emph{\bibinfo{title}{{Formal Transformation from
  Sequence Diagrams to Coloured Petri Nets}}}.
\newblock {\sl \bibinfo{journal}{2010 Asia Pacific Software Engineering
  Conference}}, pp. \bibinfo{pages}{216--225}, \doi{10.1109/APSEC.2010.33}.

\bibitemdeclare{article}{byg2009b}
\bibitem{byg2009b}
\bibinfo{author}{Joakim \surnamestart Byg\surnameend} \&
  \bibinfo{author}{Kenneth~Yrke \surnamestart J{\o}rgensen\surnameend}
  (\bibinfo{year}{2009}): \emph{\bibinfo{title}{{An Efficient Translation of
  Timed-Arc Petri Nets to Networks of Timed Automata}}}.
\newblock {\sl \bibinfo{journal}{Formal Methods and Software Engineering}}
  \bibinfo{volume}{5885}(\bibinfo{number}{1}), pp. \bibinfo{pages}{698--716},
  \doi{10.1007/978-3-642-10373-5\_36}.

\bibitemdeclare{inproceedings}{Byg2009}
\bibitem{Byg2009}
\bibinfo{author}{Joakim \surnamestart Byg\surnameend},
  \bibinfo{author}{Kenneth~Yrke \surnamestart J{\o}rgensen\surnameend} \&
  \bibinfo{author}{Jir\'{\i} \surnamestart Srba\surnameend}
  (\bibinfo{year}{2009}): \emph{\bibinfo{title}{{TAPAAL: Editor, simulator and
  verifier of timed-arc Petri nets}}}.
\newblock In: {\sl \bibinfo{booktitle}{Proceedings of the 7th International
  Symposium on Automated Technology for Verification and Analysis}}, pp.
  \bibinfo{pages}{84--89}, \doi{10.1007/978-3-642-04761-9\_7}.

\bibitemdeclare{techreport}{Damm2011}
\bibitem{Damm2011}
\bibinfo{author}{Werner \surnamestart Damm\surnameend},
  \bibinfo{author}{Andreas \surnamestart Baumgart\surnameend},
  \bibinfo{author}{Eckard \surnamestart B\"{o}de\surnameend},
  \bibinfo{author}{Matthias \surnamestart B\"{u}ker\surnameend},
  \bibinfo{author}{Tayfun \surnamestart Gezgin\surnameend},
  \bibinfo{author}{Stefan \surnamestart Henkler\surnameend},
  \bibinfo{author}{Hardi \surnamestart Hungar\surnameend},
  \bibinfo{author}{Bernhard \surnamestart Josko\surnameend},
  \bibinfo{author}{Markus \surnamestart Oertel\surnameend},
  \bibinfo{author}{Thomas \surnamestart Peikenkamp\surnameend},
  \bibinfo{author}{Philipp \surnamestart Reinkemeier\surnameend},
  \bibinfo{author}{Ingo \surnamestart Stierand\surnameend} \&
  \bibinfo{author}{Raphael \surnamestart Weber\surnameend}
  (\bibinfo{year}{2011}): \emph{\bibinfo{title}{{Architecture Modeling}}}.
\newblock \bibinfo{type}{Technical Report}, \bibinfo{institution}{OFFIS},
  \bibinfo{address}{Oldenburg}.

\bibitemdeclare{inproceedings}{Eichner2005}
\bibitem{Eichner2005}
\bibinfo{author}{Christoph \surnamestart Eichner\surnameend},
  \bibinfo{author}{Hans \surnamestart Fleischhack\surnameend} \&
  \bibinfo{author}{Roland \surnamestart Meyer\surnameend}
  (\bibinfo{year}{2005}): \emph{\bibinfo{title}{{Compositional semantics for
  UML 2.0 sequence diagrams using Petri nets}}}.
\newblock In: {\sl \bibinfo{booktitle}{In 12th Int. SDL Forum, volume 3530 of
  LNCS}}, pp. \bibinfo{pages}{133--148}, \doi{10.1007/11506843\_9}.

\bibitemdeclare{article}{Hanisch1993}
\bibitem{Hanisch1993}
\bibinfo{author}{HM~\surnamestart Hanisch\surnameend} (\bibinfo{year}{1993}):
  \emph{\bibinfo{title}{{Analysis of place/transition nets with timed arcs and
  its application to batch process control}}}.
\newblock {\sl \bibinfo{journal}{Application and Theory of Petri Nets 1993}},
  pp. \bibinfo{pages}{282--299}, \doi{10.1007/3-540-56863-8\_52}.

\bibitemdeclare{techreport}{ISO26262}
\bibitem{ISO26262}
\bibinfo{author}{\surnamestart {ISO}\surnameend} (\bibinfo{year}{2009}):
  \emph{\bibinfo{title}{{ISO/DIS 26262-1 - Road vehicles "Functional safety"
  Part 1 Glossary}}}.
\newblock \bibinfo{type}{Technical Report}.

\bibitemdeclare{inproceedings}{Li2004}
\bibitem{Li2004}
\bibinfo{author}{Xiaoshan \surnamestart Li\surnameend},
  \bibinfo{author}{Zhiming \surnamestart Liu\surnameend} \&
  \bibinfo{author}{He~\surnamestart Jifeng\surnameend} (\bibinfo{year}{2004}):
  \emph{\bibinfo{title}{{A formal semantics of UML sequence diagram}}}.
\newblock In: {\sl \bibinfo{booktitle}{Australian Software Engineering
  Conference Proceedings}}, \bibinfo{volume}{292}, pp.
  \bibinfo{pages}{168--177}, \doi{10.1109/ASWEC.2004.1290469}.

\bibitemdeclare{inproceedings}{Linzhang2004}
\bibitem{Linzhang2004}
\bibinfo{author}{W.~\surnamestart Linzhang\surnameend},
  \bibinfo{author}{Y.~\surnamestart Jiesong\surnameend},
  \bibinfo{author}{Y.~\surnamestart Xiaofeng\surnameend},
  \bibinfo{author}{H.~\surnamestart Jun\surnameend},
  \bibinfo{author}{L.~\surnamestart Xuandong\surnameend} \&
  \bibinfo{author}{Z.~\surnamestart Guoliang\surnameend}
  (\bibinfo{year}{2004}): \emph{\bibinfo{title}{{Generating test cases from UML
  activity diagram based on gray-box method}}}.
\newblock In: {\sl \bibinfo{booktitle}{Software Engineering Conference, 2004.
  11th Asia-Pacific}}, \bibinfo{volume}{60233020}, pp.
  \bibinfo{pages}{284--291}, \doi{10.1109/APSEC.2004.55}.

\bibitemdeclare{article}{Micskei2010}
\bibitem{Micskei2010}
\bibinfo{author}{Zolt\'{a}n \surnamestart Micskei\surnameend} \&
  \bibinfo{author}{H\'{e}l\`{e}ne \surnamestart Waeselynck\surnameend}
  (\bibinfo{year}{2010}): \emph{\bibinfo{title}{{The many meanings of UML 2
  Sequence Diagrams: a survey}}}.
\newblock {\sl \bibinfo{journal}{Software \& Systems Modeling}}
  \bibinfo{volume}{10}(\bibinfo{number}{4}), pp. \bibinfo{pages}{489--514},
  \doi{10.1007/s10270-010-0157-9}.

\bibitemdeclare{techreport}{uml2}
\bibitem{uml2}
\bibinfo{author}{\surnamestart OMG\surnameend} (\bibinfo{year}{2010}):
  \emph{\bibinfo{title}{{OMG Unified Modeling Language TM (OMG UML),
  Superstructure v2. 3 . 2010}}}.
\newblock \bibinfo{type}{Technical Report} \bibinfo{number}{May}.

\bibitemdeclare{phdthesis}{Sieverding2011}
\bibitem{Sieverding2011}
\bibinfo{author}{Sven \surnamestart Sieverding\surnameend}
  (\bibinfo{year}{2011}): \emph{\bibinfo{title}{{Sequenzdiagrammbasierte Test-
  und Analysemethoden von AUTOSAR-Softwarekomponenten ( SWCs )}}}.
\newblock \bibinfo{type}{Master thesis}, \bibinfo{school}{Oldenburg}.

\bibitemdeclare{article}{Sokenou2006}
\bibitem{Sokenou2006}
\bibinfo{author}{Dehla \surnamestart Sokenou\surnameend}
  (\bibinfo{year}{2006}): \emph{\bibinfo{title}{{Generating test sequences from
  UML sequence diagrams and state diagrams}}}.
\newblock {\sl \bibinfo{journal}{Informatik 2006: Informatik f\"{u}r Menschen}}
  \bibinfo{volume}{2}(\bibinfo{number}{94}), pp. \bibinfo{pages}{236--240}.

\bibitemdeclare{inproceedings}{Srba2005}
\bibitem{Srba2005}
\bibinfo{author}{Jir\'{\i} \surnamestart Srba\surnameend}
  (\bibinfo{year}{2005}): \emph{\bibinfo{title}{{Timed-arc Petri nets vs.
  networks of timed automata}}}.
\newblock In: {\sl \bibinfo{booktitle}{Applications and Theory of Petri Nets}},
  pp. \bibinfo{pages}{1273--1278}, \doi{10.1007/11494744\_22}.

\bibitemdeclare{article}{srba2008comparing}
\bibitem{srba2008comparing}
\bibinfo{author}{Jir\'{\i} \surnamestart Srba\surnameend}
  (\bibinfo{year}{2008}): \emph{\bibinfo{title}{Comparing the expressiveness of
  timed automata and timed extensions of Petri nets}}.
\newblock {\sl \bibinfo{journal}{Formal Modeling and Analysis of Timed
  Systems}}, pp. \bibinfo{pages}{15--32}, \doi{10.1007/978-3-540-85778-5\_3}.

\end{thebibliography}
